\documentclass[prb,preprint,showpacs,preprintnumbers,amsmath,amssymb,unsortedaddress]{revtex4-1}

\usepackage{graphicx}
\usepackage{dcolumn}
\usepackage{bm}
\usepackage{amsthm}
\usepackage{amsmath}
\usepackage{amssymb}

\begin{document}

\title{Magnetic Guinier Law}

\author{Andreas~Michels}\email[Corresponding author: ]{andreas.michels@uni.lu}
\author{Artem~Malyeyev}
\author{Ivan~Titov}
\author{Dirk~Honecker}
\affiliation{Physics and Materials Science Research Unit, University of Luxembourg, 162A~Avenue de la Fa\"iencerie, L-1511 Luxembourg, Grand Duchy of Luxembourg}
\author{Robert~Cubitt}
\affiliation{Institut Laue-Langevin, 71~avenue des Martyrs, F-38042 Grenoble, France}
\author{Elizabeth~Blackburn}
\affiliation{Division of Synchrotron Radiation Research, Department of Physics, Lund University, SE-22100 Lund, Sweden}
\author{Kiyonori~Suzuki}
\affiliation{Department of Materials Science and Engineering, Monash University, Clayton, Victoria~3800, Australia}

\keywords{small-angle neutron scattering, Guinier law, magnetic materials, micromagnetics}

\begin{abstract}
Small-angle scattering of x-rays and neutrons is a routine method for the determination of nanoparticle sizes. The so-called Guinier law represents the low-$q$ approximation for the small-angle scattering curve from an assembly of particles. The Guinier law has originally been derived for \textit{nonmagnetic} particle-matrix-type systems, and it is successfully employed for the estimation of particle sizes in various scientific domains (e.g., soft matter physics, biology, colloidal chemistry, materials science). An important prerequisite for it to apply is the presence of a \textit{discontinuous} interface separating particles and matrix. Here, we introduce the Guinier law for the case of \textit{magnetic} small-angle neutron scattering (SANS) and experimentally demonstrate its applicability for the example of nanocrystalline cobalt. It is well-known that the magnetic microstructure of nanocrystalline ferromagnets is highly nonuniform on the nanometer length scale and characterized by a spectrum of \textit{continuously} varying long-wavelength magnetization fluctuations, i.e., these systems do not manifest sharp interfaces in their magnetization profile. The magnetic Guinier radius depends on the applied magnetic field, on the magnetic interactions (exchange, magnetostatics), and on the magnetic anisotropy-field radius, which characterizes the size over which the magnetic anisotropy field is coherently aligned into the same direction. In contrast to the nonmagnetic conventional Guinier law, the magnetic version can be applied to fully dense random-anisotropy-type ferromagnets.
\end{abstract}

\maketitle\

\section{Introduction}

The determination of particle sizes is one of the most important daily tasks in many branches of the natural sciences. While particle sizes in the micrometer regime and above can be conveniently determined using e.g.\ optical microscopy, the size of nanoparticles (with $D \sim 1-100 \, \mathrm{nm}$) requires scanning and/or transmission electron microscopy or other scattering methods such as x-ray or neutron scattering. While the former techniques inherently suffer from a low statistics, the latter ones have the advantage of providing statistically-averaged information over a large of number of particles. Small-angle scattering, using either x-rays or neutrons, is one of the most popular methods for analyzing structures on this mesoscopic length scale, embracing a broad range of research topics from condensed-matter and soft-matter physics, physical chemistry, biology, and materials science~\cite{svergun2013}.

The well-known Guinier law describes the elastic small-angle scattering of x-rays and neutrons near the origin of reciprocal space~\cite{fournet}: when the scattering is from a dilute and monodisperse set of objects (particles) with sharp interfaces, then the macroscopic differential scattering cross section $d \Sigma / d \Omega$ in the limit of low momentum transfers $q < 1.3/R_G$ can be expressed as~\cite{porod,feigin}:
\begin{eqnarray}
\label{guinierlaw}
\frac{d \Sigma}{d \Omega}(q) \cong \frac{d \Sigma}{d \Omega}(q = 0) \, e^{- \frac{q^2 R_G^2}{3}} ,
\end{eqnarray}
where the forward scattering cross section $\frac{d \Sigma}{d \Omega}(0)$ is proportional to the squared total excess scattering length of the particle, and $R_G$ denotes the particle's radius of gyration. Equation~(\ref{guinierlaw}) is valid for arbitrary particle shapes. From a Guinier plot, $\ln(d \Sigma / d \Omega)$ \textit{vs.}\ $q^2$, one can determine $R_G$, which is related to the particle size, e.g., $R_G^2 = \frac{3}{5} R^2$ for a sphere of radius $R$. The Guinier law is of outstanding importance for the analysis of small-angle scattering data, particularly at the first stage of the data analysis.

From the foregoing discussion it is clear that the Guinier law has been derived for \textit{nonmagnetic} particle-matrix-type assemblies in the context of the early theoretical developments of the technique of small-angle x-ray scattering~\cite{fournet}. Therefore, its application to magnetic materials, which is the subject of the present paper, should be considered with special care; for instance, the Guinier law is certainly applicable to systems consisting of saturated and homogeneous magnetic particles in a nonmagnetic and homogeneous matrix or, likewise, to pores in a saturated matrix. In this context we refer to the paper by Burke~\cite{burke1981} who investigated the influence of magnetic shape anisotropy on the Guinier law of fine ferromagnetic \textit{single-domain} particles. By contrast, when the sample is \textit{inhomogeneously} magnetized on the nanometer length scale, i.e., when the magnitude and orientation of the magnetization vector field $\mathbf{M}$ varies continuously with the position $\mathbf{r}$ inside the material, then a central assumption underlying the Guinier law---namely that of domains (particles) separated by discontinuous interfaces from the matrix---is violated. Equation~(\ref{guinierlaw}), with a constant and field-independent $R_G$, does not then describe the low-$q$ region of the magnetic SANS cross section. Intuitively, it may be clear from the previous considerations that an effective magnetic Guinier radius is expected to depend on the applied magnetic field as well as on the magnetic interactions (e.g., exchange, anisotropy, magnetostatics). In the following we derive the magnetic Guinier law and provide an analysis of experimental SANS data of nanocrystalline Co.

The paper is organized as follows:~Sec.~\ref{exp} furnishes the details of the SANS experiment; Secs.~\ref{mumag} and \ref{glaw} introduce the unpolarized SANS cross section, the theoretical background in terms of micromagnetic theory, and the magnetic Guinier law; Sec.~\ref{results} presents and discusses the experimental results of the magnetic Guinier analysis on nanocrystalline Co; Sec.~\ref{conclusion} summarizes the main results of this study. In the Supplemental Material to this paper~\cite{smguinier2019} the two and one-dimensional total SANS cross sections and a graphical representation of the relative error of the magnetic Guinier approximation are featured.

\section{Experimental}
\label{exp}

\begin{figure}[tb!]
\centering
\resizebox{0.85\columnwidth}{!}{\includegraphics{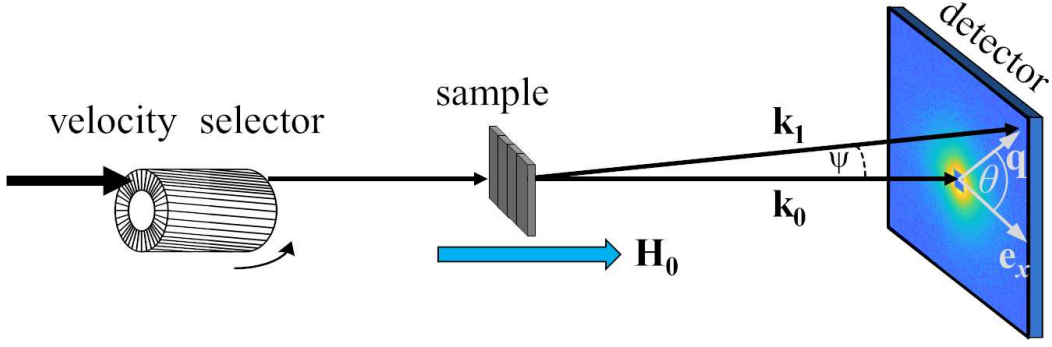}}
\caption{Sketch of the neutron setup. The external magnetic field $\mathbf{H}_0 \parallel \mathbf{e}_z$ is applied parallel to the wave vector $\mathbf{k}_0$ of the incident neutrons. In the small-angle approximation the momentum-transfer or scattering vector $\mathbf{q} = \mathbf{k}_1 - \mathbf{k}_0$ varies in the plane perpendicular to $\mathbf{k}_0$, i.e., $\mathbf{q} \cong \{q_x, q_y, 0 \} = q \{\cos\theta, \sin\theta, 0 \}$. The magnitude of $\mathbf{q}$ for elastic scattering is given by $q = \frac{4\pi}{\lambda} \sin(\psi/2)$, where $\lambda$ denotes the mean neutron wavelength (selected by the velocity selector) and $\psi$ is the scattering angle. The angle $\theta$ specifies the orientation of $\mathbf{q}$ on the two-dimensional detector.}
\label{sanssetup}
\end{figure}

The SANS experiment was conducted at $300 \, \mathrm{K}$ at the instrument D11 at the Institut Laue-Langevin, Grenoble. We used unpolarized incident neutrons with a mean wavelength of $\lambda = 6.0 \, \mathrm{\AA}$ and a bandwidth of $\Delta \lambda / \lambda = 10 \, \%$ (FWHM). The instrument offers access to a low $q$-range of $0.016 \, \mathrm{nm}^{-1} \lesssim q \lesssim 0.2 \, \mathrm{nm}^{-1}$ with the two-dimensional position-sensitive detector placed at a distance of $38.5 \, \mathrm{m}$ from the sample position. The external magnetic field $\mathbf{H}_0$ (with $\mu_0 H_0^{\mathrm{max}} = 16.5 \, \mathrm{T}$) was applied parallel to the wave vector $\mathbf{k}_0$ of the incoming neutron beam (see Fig.~\ref{sanssetup} for a sketch of the neutron setup).

The nanocrystalline Co sample under study was synthesized by means of pulsed electrodeposition. We emphasize that this particular sample has been extensively studied in the past using magnetometry, wide-angle X-ray diffraction, and unpolarized and spin-polarized SANS (e.g., \cite{michels00b,jprb2001,michels03prl,michels2011jpcm,michels2014review,mettus2015}). It is also important to note that it is a fully dense polycrystalline bulk metal with a nanometer grain size (average crystallite size:~$D = 9.5 \pm 3.0 \, \mathrm{nm}$~\cite{jprb2001}), not nanoparticles in a matrix. The SANS sample consisted of a single circular disk. Based on the thickness ($80 \, \mu{\mathrm{m}}$) and the diameter ($2 \, {\mathrm{cm}}$) of the disk, we computed a demagnetizing factor of $N \cong 0.994$ for the case that $\mathbf{H}_0$ is parallel to the surface normal of the sample~\cite{osborn}, in agreement with the $\mathbf{k}_0 \parallel \mathbf{H}_0$ scattering geometry of the SANS experiment. Using the value of the saturation magnetization of Co, $\mu_0 M_s = 1.80 \, \mathrm{T}$ ($\hat{=} 1434 \, \mathrm{kA/m}$), this results in a demagnetizing field of $\mu_0 N M_s \cong 1.789 \, \mathrm{T}$ for a fully saturated sample. In the following all the reported field values are corrected for demagnetizing effects. To reduce the influence of inhomogeneous demagnetizing fields at the outer perimeter of the circular sample, the neutron beam was collimated to a diameter of $0.8 \, {\mathrm{cm}}$. The neutron transmission was larger than $90 \, \%$ in all measurements, indicating a negligible influence of multiple scattering.

\section{SANS cross section and micromagnetic theory}
\label{mumag}

When the external magnetic field $\mathbf{H}_0 \parallel \mathbf{e}_z$ is applied parallel to the wave vector $\mathbf{k}_0$ of the incoming neutron beam (Fig.~\ref{sanssetup}), the unpolarized elastic differential SANS cross section $d \Sigma / d \Omega$ at momentum-transfer vector $\mathbf{q}$ equals~\cite{rmp2019}:
\begin{eqnarray}
\label{sigmasanspara}
\frac{d \Sigma}{d \Omega}(\mathbf{q}) = \frac{8 \pi^3}{V} b_H^2 \left( \frac{|\widetilde{N}|^2}{b_H^2} + |\widetilde{M}_x|^2 \sin^2\theta + |\widetilde{M}_y|^2 \cos^2\theta + |\widetilde{M}_z|^2 \right. \nonumber \\ \left. - (\widetilde{M}_x \widetilde{M}_y^{\ast} + \widetilde{M}_x^{\ast} \widetilde{M}_y) \sin\theta \cos\theta \right) ,
\end{eqnarray}
where $V$ is the scattering volume, the constant $b_H = 2.91 \times 10^{8} \mathrm{A^{-1} m^{-1}}$ relates the atomic magnetic moment $\mu_a$ to the atomic magnetic scattering length $b_m \cong b_H \mu_a$ (in small-angle approximation), $\widetilde{N}(\mathbf{q})$ and $\mathbf{\widetilde{M}}(\mathbf{q}) = \{ \widetilde{M}_x(\mathbf{q}), \widetilde{M}_y(\mathbf{q}), \widetilde{M}_z(\mathbf{q}) \}$ represent, respectively, the Fourier transforms of the nuclear scattering length density $N(\mathbf{r})$ and of the magnetization $\mathbf{M}(\mathbf{r}) = \{ M_x(\mathbf{r}), M_y(\mathbf{r}), M_z(\mathbf{r}) \}$, the superscript ``$\, ^{\ast} \,$'' refers to the complex-conjugated quantity, and $\theta$ denotes the angle between $\mathbf{q}$ and $\mathbf{e}_x$. Note that in the small-angle approximation the component of $\mathbf{q}$ along the incident beam ($\mathbf{k}_0 \parallel \mathbf{e}_z$) is negligible as compared to the other two components, such that $\mathbf{q} \cong \{ q_x, q_y, 0 \}$. This emphasizes the fact that SANS predominantly probes correlations in the plane perpendicular to $\mathbf{k}_0$.

The further analysis of the magnetic SANS cross section Eq.~(\ref{sigmasanspara}) requires expressions for the magnetization Fourier amplitudes $\widetilde{M}_{x,y,z}$. In Refs.~\onlinecite{michels2013,michelsPRB2016} a quite general theory of magnetic SANS based on the continuum theory of micromagnetics has been developed. In the following we sketch the main ideas of the micromagnetic SANS theory in order to achieve a self-contained presentation. The approach considers two origins of spin misalignment:~(i)~Spatial nanometer-scale variations in the orientation and/or magnitude of the magnetic anisotropy field $\mathbf{H}_p(\mathbf{r})$ (e.g., at a grain boundary in a single-phase nanocrystalline ferromagnet). Such anisotropy-field fluctuations give rise to torques on the magnetization~$\mathbf{M}$ and result in a concomitant deviation of $\mathbf{M}$ from the mean magnetization direction (given by a large applied field). (ii)~Spatial variations of the saturation magnetization $M_s(\mathbf{r})$ give rise to local magnetostatic stray fields (e.g., at a particle-matrix interphase in a nanocomposite), which also result in a magnetic SANS contrast. This scenario is adapted to the inhomogeneous magnetic microstructure which is found in many polycrystalline magnets.

The micromagnetic theory takes into account the isotropic and symmetric exchange interaction, magnetic anisotropy, as well as the Zeeman and the magnetodipolar interaction energies. As detailed in the pertinent textbooks~\cite{brown,aharonibook,kronfahn03,mumagcoll}, variational calculus leads to a set of nonlinear partial differential equations for the equilibrium magnetization configuration $\mathbf{M}(\mathbf{r})$. For the static case, the equations of micromagnetics (so-called Brown's equations) can be conveniently expressed as a balance-of-torques equation:
\begin{equation}
\label{torque}
\mathbf{M}(\mathbf{r}) \times \mathbf{H}_{\mathrm{eff}}(\mathbf{r}) = 0 .
\end{equation}
Equation~(\ref{torque}) expresses the fact that at static equilibrium the torque on the magnetization $\mathbf{M}(\mathbf{r})$ due to an effective magnetic field $\mathbf{H}_{\mathrm{eff}}(\mathbf{r})$ vanishes at each point $\mathbf{r}$ inside the material. The effective field is obtained as:
\begin{eqnarray}
\label{heff}
\mathbf{H}_{\mathrm{eff}}(\mathbf{r}) = \mathbf{H}_{\mathrm{ex}}(\mathbf{r}) + \mathbf{H}_p(\mathbf{r}) + \mathbf{H}_0 + \mathbf{H}_d(\mathbf{r}) ,
\end{eqnarray}
where $\mathbf{H}_{\mathrm{ex}}(\mathbf{r}) = l_M^2 \Delta \mathbf{M}(\mathbf{r})$ represents the exchange field (with $\Delta$ the Laplace operator), $\mathbf{H}_p(\mathbf{r})$ is the magnetic anisotropy field, $\mathbf{H}_0$ is a uniform applied magnetic field, and $\mathbf{H}_d(\mathbf{r})$ denotes the magnetostatic or magnetodipolar interaction field. The micromagnetic exchange length $l_M = \sqrt{2 A / (\mu_0 M_s^2)}$ is of the order of a few nanometers for many magnetic materials ($l_M \sim 3-10 \, \mathrm{nm}$~\cite{kronfahn03}), $A$ is the exchange-stiffness constant, $M_s$ is the saturation magnetization, and $\mu_0$ denotes the permeability of free space. Then, in the approach-to-saturation regime, the micromagnetic equations can be linearized and closed-form expressions for the magnetization Fourier components $\widetilde{M}_x(\mathbf{q})$ and $\widetilde{M}_y(\mathbf{q})$ can be obtained (see Refs.~\onlinecite{michels2013,michelsPRB2016} for details).

Using the results for $\widetilde{M}_x$ and $\widetilde{M}_y$ the unpolarized elastic SANS cross section $d \Sigma / d \Omega$ in the parallel scattering geometry [Eq.~(\ref{sigmasanspara})] can be expressed in compact form as:
\begin{eqnarray}
\frac{d \Sigma}{d \Omega}(q,H_i) = \frac{d \Sigma_{\mathrm{res}}}{d \Omega}(q) + S_H(q) R_H(q, H_i) ,
\label{dsigmatot}
\end{eqnarray}
where the (nuclear and magnetic) so-called residual SANS cross section
\begin{eqnarray}
\frac{d \Sigma_{\mathrm{res}}}{d \Omega} = \frac{8 \pi^3}{V} (|\widetilde{N}|^2 + b^2_H |\widetilde{M}_s|^2)
\label{dsigmaresidual}
\end{eqnarray}
is measured at complete magnetic saturation ($|\widetilde{M}_z|^2 = |\widetilde{M}_s|^2$), and the remaining spin-misalignment SANS cross section
\begin{eqnarray}
\frac{d \Sigma_{SM}}{d \Omega} &=& \frac{8 \pi^3}{V} b_H^2 \left( |\widetilde{M}_x|^2 \sin^2\theta + |\widetilde{M}_y|^2 \cos^2\theta - (\widetilde{M}_x \widetilde{M}_y^{\ast} + \widetilde{M}_x^{\ast} \widetilde{M}_y) \sin\theta \cos\theta \right) \nonumber \\ &=& S_H(q) R_H(q, H_i) ,
\label{dsigmasm}
\end{eqnarray}
describes the purely magnetic small-angle scattering due to the misaligned spins with related Fourier amplitudes $\widetilde{M}_x(\mathbf{q})$ and $\widetilde{M}_y(\mathbf{q})$ [compare Eq.~(\ref{sigmasanspara})]. Since the magnetic Guinier law is related to the spin-misalignment scattering, it is necessary to separate the total $d \Sigma / d \Omega$ into $d \Sigma_{\mathrm{res}} / d \Omega$ and $d \Sigma_{SM} / d \Omega$ [Eq.~(\ref{dsigmatot})]. In the analysis of experimental data, $d \Sigma_{\mathrm{res}} / d \Omega$ can be measured at a saturating applied field and subtracted from the $d \Sigma / d \Omega$ at lower fields to obtain the field-dependent $d \Sigma_{SM} / d \Omega$.

The quantity $S_H$ denotes the anisotropy-field scattering function, which is proportional to the magnitude square of the Fourier transform $\mathbf{\widetilde{H}}_p(\mathbf{q})$ of the magnetic anisotropy field $\mathbf{H}_p(\mathbf{r})$, i.e., $S_H \propto \mathbf{\widetilde{H}}^2_p(\mathbf{q})$. This function contains information on the strength and spatial structure of the magnetic anisotropy field. In the approach-to-saturation regime, which is the validity range of the SANS theory, $S_H$ is independent of the applied magnetic field. We further note that for a statistically isotropic material $S_H$ depends only on the magnitude $q$ of the scattering vector $\mathbf{q}$, not on its orientation (see below). The \textit{dimensionless} micromagnetic response function $R_H$ depends on $q$ as well as on the internal magnetic field $H_i = H_0 - N M_s$, where $N$ denotes the demagnetizing factor. More specifically ($\mathbf{k}_0 \parallel \mathbf{H}_0$),
\begin{eqnarray}
\label{resppara}
R_H(q, H_i) = \frac{p^2(q, H_i)}{2} ,
\end{eqnarray}
where the dimensionless function
\begin{eqnarray}
\label{pdef}
p(q, H_i) = \frac{M_s}{H_{\mathrm{eff}}(q, H_i)} = \frac{M_s}{H_i (1 + l_H^2 q^2)}
\end{eqnarray}
depends on the effective magnetic field $H_{\mathrm{eff}}(q, H_i)$ [not to be confused with $\mathbf{H}_{\mathrm{eff}}(\mathbf{r})$ in Eq.~(\ref{torque})], and on the micromagnetic exchange-length
\begin{eqnarray}
\label{lhdef}
l_H(H_i) = \sqrt{\frac{2 A}{\mu_0 M_s H_i}} .
\end{eqnarray}
The quantity $l_H$ characterizes the field-dependent size of perturbed regions around microstructural defects, and as we will see below it is this quantity which renders the magnetic Guinier radius field dependent. By inserting typical values for the material parameters of Co ($A = 2.8 \times 10^{-11} \, \mathrm{J/m}$ and $\mu_0 M_s = 1.80 \, \mathrm{T}$~\cite{michels08rop}), it is seen that the exchange length $l_H$ varies between about $200-2 \, \mathrm{nm}$ when the internal field is changed between $0.001 - 10 \, \mathrm{T}$. This length scale falls well into the resolution regime of the SANS technique.

\section{Magnetic Guinier law}
\label{glaw}

In order to derive a Guinier expression for magnetic SANS, analogous to Eq.~(\ref{guinierlaw}), we look in the following for the low-$q$ behavior of the spin-misalignment SANS cross section $d \Sigma_{SM} / d \Omega = S_H(q) R_H(q, H_i)$ [Eq.~(\ref{dsigmasm})]. The sample volume which is probed by the neutrons typically contains many defects (e.g., crystallites separated by grain boundaries), each one having a different orientation and/or magnitude of the magnetic anisotropy field. To obtain a low-$q$ approximation for $S_H \propto \mathbf{\widetilde{H}}^2_p(\mathbf{q})$, we make the assumption that the total magnetic anisotropy field of the sample, $\mathbf{H}_p(\mathbf{r})$, is the sum of the anisotropy fields of the individual defects ``$i$''~\cite{jnist,jprb2001}, i.e.,
\begin{eqnarray}
\label{Hpjsuperpo}
\mathbf{H}_p(\mathbf{r}) = \sum\limits_{i=1}^N \mathbf{H}_{p,i}(\mathbf{r}) .
\end{eqnarray}
This decomposition also applies to the Fourier transform $\mathbf{\widetilde{H}}_p(\mathbf{q})$ of $\mathbf{H}_p(\mathbf{r})$, i.e.,
\begin{eqnarray}
\label{hjsuperpo}
\mathbf{\widetilde{H}}_p(\mathbf{q}) = \sum\limits_{i=1}^N \mathbf{\widetilde{H}}_{p,i}(\mathbf{q}) ,
\end{eqnarray}
so that
\begin{eqnarray}
\label{hjsuperposquare}
\mathbf{\widetilde{H}}^2_p = \sum\limits_{i=1}^N \mathbf{\widetilde{H}}^2_{p,i} + \sum\limits_{i \neq j}^N \mathbf{\widetilde{H}}_{p,i} \cdot \mathbf{\widetilde{H}}_{p,j} ,
\end{eqnarray}
where we have assumed that the $\mathbf{\widetilde{H}}_{p,i}$ are real-valued quantities. If the $\mathbf{\widetilde{H}}_{p,i}$ of the individual defects are statistically uncorrelated (random anisotropy), then terms $\mathbf{\widetilde{H}}_{p,i} \cdot \mathbf{\widetilde{H}}_{p,j}$ with $i \neq j$ take on both signs with equal probability. Consequently, the sum over these terms vanishes, and
\begin{eqnarray}
\label{hjsuperposquare}
\mathbf{\widetilde{H}}^2_p(\mathbf{q}) = \sum\limits_{i=1}^N \mathbf{\widetilde{H}}^2_{p,i}(\mathbf{q}) .
\end{eqnarray}
Equation~(\ref{hjsuperposquare}) suggests that $\mathbf{\widetilde{H}}^2_p$, and hence $S_H \propto \mathbf{\widetilde{H}}^2_p$, can be computed for an arbitrary arrangement of defects once the solution for the single-defect case $\mathbf{\widetilde{H}}_{p,i}(\mathbf{q})$ is known. This can e.g.\ be accomplished for an idealized nanocrystalline ferromagnet, where the crystallites (acting as a ``magnetic defects'') have random crystallographic orientation and where the anisotropy field arises exclusively from the magnetocrystalline anisotropy. Because each grain is a single crystal, the anisotropy field in the grain is a constant vector, i.e., $\mathbf{H}_{p,i} \neq \mathbf{H}_{p,i}(\mathbf{r})$, and the anisotropy field Fourier amplitude is obtained by the following form-factor integral~\cite{jprb2001}:
\begin{eqnarray}
\label{hjsinglegrain}
\mathbf{\widetilde{H}}_{p,i}(\mathbf{q}) = \frac{\mathbf{H}_{p,i}}{(2\pi)^{3/2}} \int_{V_{p,i}} e^{- i \mathbf{q} \cdot \mathbf{r}} d^3r ,
\end{eqnarray}
where the integral extends over the volume of grain~$i$. For the example of a spherical grain shape ($V_{p,i} = \frac{4\pi}{3} R_i^3$), we obtain the well-known result that
\begin{eqnarray}
\label{hjsinglegrainsphere}
\mathbf{\widetilde{H}}_{p,i}(\mathbf{q}) = \mathbf{\widetilde{H}}_{p,i}(q R_i) = \frac{\mathbf{H}_{p,i}}{(2\pi)^{3/2}} 3 V_{p,i} \frac{j_1(q R_i)}{q R_i} ,
\end{eqnarray}
where $j_1(z)$ denotes the spherical Bessel function of the first order.

The square of Eq.~(\ref{hjsinglegrainsphere}) is identical, except for the prefactor, to the nuclear SANS cross section of an array of noninterfering spherical particles, and general asymptotic results at small and large $q$ are therefore immediately transferable; in particular, the Guinier approximation relates $S_H(q) \propto \widetilde{H}^2_p(q)$ at small scattering vectors to the radius of gyration $R_{GH}$ of the magnetic anisotropy field, according to~\cite{jprb2001}
\begin{eqnarray}
\label{guinierlawsh}
S_H(q) \cong S_H(0) \, e^{- \frac{q^2 R_{GH}^2}{3}} .
\end{eqnarray}
Similar to nuclear SANS and SAXS, where $R_G$ is a measure for the particle size, $R_{GH}$ deduced from $S_H$ may be seen as a measure for the size of regions over which the magnetic anisotropy field $\mathbf{H}_p(\mathbf{r})$ is homogeneous. For the special case of an idealized nanocrystalline ferromagnet (random anisotropy and magnetocrystalline anisotropy only), $R_{GH}$ is closely related to the crystallite size.

Equation~(\ref{guinierlawsh}) can be combined with the corresponding small-$q$ result for the response function [Eq.~(\ref{resppara})]. Taylor expansion of $R_H$ around $q = 0$ yields:
\begin{eqnarray}
\label{rhtaylorpara}
R_H(q, H_i) = \frac{p^2(q, H_i)}{2} \cong \frac{p_0^2}{2} \left( 1 - 2 l_H^2 q^2 \right) \cong \frac{p_0^2}{2} e^{- \frac{q^2 6 l_H^2}{3}} ,
\end{eqnarray}
where $p_0 = p(q = 0) = M_s/H_i$ [compare Eq.~(\ref{pdef})]. Inserting Eqs.~(\ref{guinierlawsh}) and (\ref{rhtaylorpara}) into $d \Sigma_{SM} / d \Omega = S_H R_H$, we have
\begin{eqnarray}
\label{sigmamtaylorall}
\frac{d \Sigma_{SM}}{d \Omega} \cong \frac{d \Sigma_{SM}}{d \Omega}(q = 0) \, e^{- \frac{q^2 R_{GSM}^2}{3}} ,
\end{eqnarray}
where 
\begin{eqnarray}
\label{refftaylorpara}
R_{GSM}^2(H_i) = R_{GH}^2 + 6 l_H^2(H_i) = R_{GH}^2 + \frac{12A}{\mu_0 M_s H_i}
\end{eqnarray}
represents the magnetic-field-dependent magnetic Guinier radius. This relation provides a means to determine the exchange constant $A$ from field-dependent SANS measurements. Note that $\frac{d \Sigma_{SM}}{d \Omega}(0) \propto p_0^2 \propto H_i^{-2}$ [compare Eq.~(\ref{pdef})]. The observation that $R_{GSM}$ depends on $R_{GH}$ and on the micromagnetic exchange length $l_H$ is a manifestation of the fact that the magnetic microstructure in real space (for which $R_{GSM}$ is representative) corresponds to the convolution of the nuclear grain microstructure ($R_{GH}$) with field-dependent micromagnetic response functions ($l_H$).

Up to now we have only discussed the magnetic Guinier approximation for the parallel scattering geometry ($\mathbf{k}_0 \parallel \mathbf{H}_0$), where $2\pi$-averaged magnetic SANS data can be used for the analysis in terms of Eq.~(\ref{sigmamtaylorall}). In the perpendicular geometry ($\mathbf{k}_0 \perp \mathbf{H}_0$) an additional scattering term $S_M R_M$, related to magnetostatic fluctuations, appears in $d \Sigma_{SM} / d \Omega$, which complicates the discussion. Two comments are in place:~(i)~Since $S_M \propto \widetilde{M}^2_{z}(\mathbf{q})$, the $S_M R_M$ contribution to $d \Sigma_{SM} / d \Omega$ can be neglected for single-phase ferromagnets, where fluctuations in the saturation magnetization $M_s$ are weak. (ii)~Inspection of the expression for the magnetostatic response function $R_M$ in the perpendicular geometry (Eq.~(29) in \cite{michels2013}) shows that this function vanishes by taking an average of the two-dimensional $d \Sigma_{SM} / d \Omega$ along $\theta = 0^{\circ}$ (or $\theta = 180^{\circ}$), while the corresponding $R_H(\theta = 0^{\circ}) = p^2$ (Eq.~(28) in \cite{michels2013}) is almost equal (besides a factor of $1/2$) to $R_H(\theta = 0^{\circ}) = p^2/2$ in the parallel geometry. In other words, these considerations imply that the magnetic Guinier law [Eq.~(\ref{sigmamtaylorall})] can also be employed to analyze ($\theta = 0^{\circ}$) sector-averaged data in the $\mathbf{k}_0 \perp \mathbf{H}_0$ geometry.

\section{Experimental results and discussion}
\label{results}

\begin{figure}[tb!]
\centering
\resizebox{1.0\columnwidth}{!}{\includegraphics{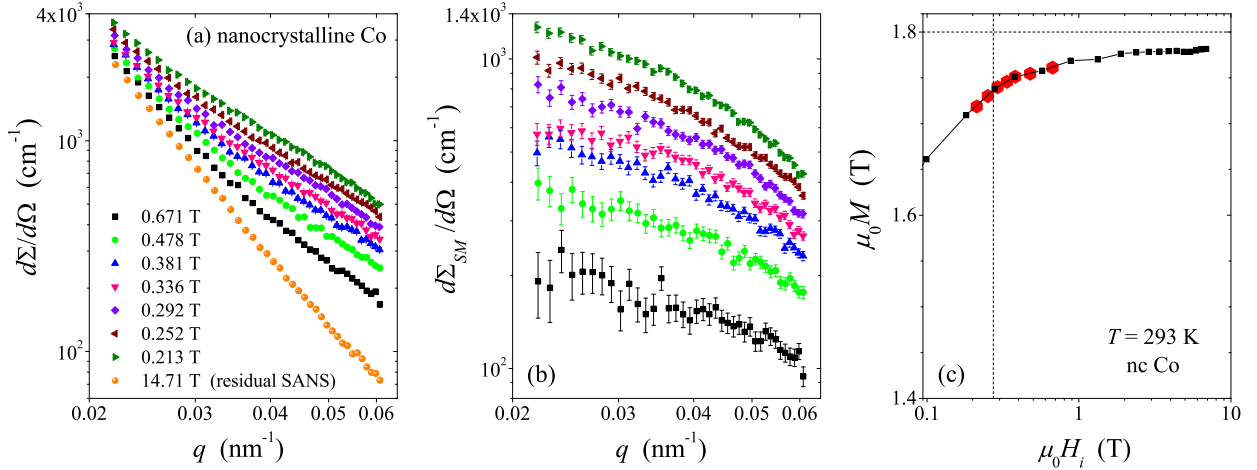}}
\caption{(a)~$2\pi$-azimuthally-averaged total nuclear and magnetic SANS cross section $d \Sigma / d \Omega$ of nanocrystalline Co \textit{vs.}\ momentum transfer $q$ at a series of internal magnetic fields (see inset) (log-log scale) ($\mathbf{k}_0 \parallel \mathbf{H}_0$). (b)~Corresponding spin-misalignment SANS cross section $d \Sigma_{SM} / d \Omega$ obtained by subtracting the $d \Sigma / d \Omega$ data at $14.71 \, \mathrm{T}$ [orange data points in (a)] from the $d \Sigma / d \Omega$ at lower fields. (c)~Magnetization curve of nanocrystalline Co (only the upper right quadrant is shown). The large red data points indicate the internal-field values where the SANS data were taken. Horizontal dashed line indicates the saturation-magnetization value of $\mu_0 M_s = 1.80 \, \mathrm{T}$. Vertical dashed line indicates the approach-to-saturation regime ($M/M_s \gtrsim 96 \, \%$).}
\label{sansdatasubtraction}
\end{figure}

As expected, the two-dimensional SANS intensity distributions of the nanocrystalline Co sample are isotropic ($\theta$-independent) at all fields investigated~\cite{commentmguinier2019} (see the Supplemental Material~\cite{smguinier2019}). By contrast, for the perpendicular scattering geometry ($\mathbf{k}_0 \perp \mathbf{H}_0$), the magnetic SANS cross section of untextured samples exhibits a variety of angular anisotropies (e.g., \cite{loeff05,bischof07,michels2014jmmm}). This supports the assumption made in the micromagnetic theory of a statistically-isotropic grain microstructure. The two-dimensional nuclear and magnetic SANS data were azimuthally-averaged over an angle of $2\pi$. To apply Eqs.~(\ref{sigmamtaylorall}) and (\ref{refftaylorpara}) to experimental $d \Sigma_{SM} / d \Omega$ data [compare Eq.~(\ref{dsigmasm})], one needs to subtract the scattering close to saturation (here:~at $14.71 \, \mathrm{T}$ internal field), corresponding to the residual SANS cross section $d \Sigma_{\mathrm{res}} / d \Omega$ [Eq.~(\ref{dsigmaresidual})], from the total $d \Sigma / d \Omega$ at lower fields [Eq.~(\ref{dsigmatot})]. The subtraction procedure along with the room-temperature magnetization curve is depicted in Fig.~\ref{sansdatasubtraction}. Besides eliminating the nuclear and the longitudinal magnetic scattering, the subtraction also removes any background scattering contribution.

By inspection of Fig.~\ref{sansdatasubtraction}(c) we see that the magnetization state of the specimen used in the SANS experiment (indicated by the large red data points) falls well into the approach-to-saturation regime, which is reached for $\mu_0 H_i \gtrsim 0.27 \, \mathrm{T}$ ($M/M_s \gtrsim 96 \, \%$, see the discussion below). The shape of $d \Sigma_{SM} / d \Omega$ is substantially different to that of $d \Sigma / d \Omega$, which is due to the subtraction of the nuclear and saturation scattering (see also the discussion in Ref.~\onlinecite{bickapl2013}). When the internal field is decreased from $0.671 \, \mathrm{T}$ to $0.213 \, \mathrm{T}$, $d \Sigma_{SM} / d \Omega$ increases strongly by a factor of $\sim 6-7$ at the smallest momentum-transfers $q$. The strong field dependence of $d \Sigma_{SM} / d \Omega$ supports the notion that scattering due to transversal spin misalignment represents by far the dominant contribution to $d \Sigma / d \Omega$ (see also Fig.~3 in \cite{michels2014review}). The experimental neutron data in Fig.~\ref{sansdatasubtraction}(b) cannot be reproduced by decomposing the cross section into a set of noninterfering single-domain particles. Careful scrutiny of Fig.~\ref{sansdatasubtraction}(b) reveals that the point with the largest curvature in $d \Sigma_{SM} / d \Omega$ evolves to larger $q$-values with increasing $H_i$, in agreement with the concomitant decrease of the exchange length $l_H$ in Eq.~(\ref{refftaylorpara}).

\begin{figure}[tb!]
\centering
\resizebox{1.0\columnwidth}{!}{\includegraphics{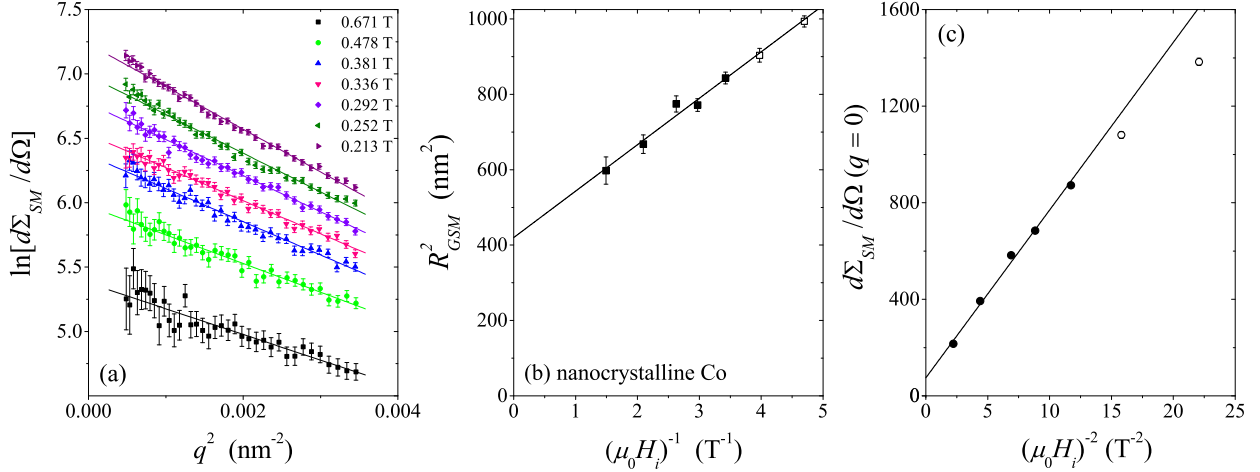}}
\caption{Magnetic Guinier analysis on nanocrystalline Co. (a)~Guinier plot $\ln[d \Sigma_{SM} / d \Omega]$ \textit{vs.} $q^2$ and fits (solid lines) to Eq.~(\ref{sigmamtaylorall}) at selected values of the internal magnetic field (see inset). (b)~Plot of $R_{GSM}^2$ \textit{vs.} $H_i^{-1}$ and fit (solid line) to Eq.~(\ref{refftaylorpara}). In the fitting routine $R_{GH}$ and $A$ were treated as adjustable parameters. (c)~Field dependence of $\frac{d \Sigma_{SM}}{d \Omega}(q = 0)$. Solid line:~$\frac{d \Sigma_{SM}}{d \Omega}(q = 0) \propto H_i^{-2}$. In~(b) and~(c) the last two data points (open symbols), corresponding to internal fields of $0.213 \, \mathrm{T}$ and $0.252 \, \mathrm{T}$, have been excluded from the fit analysis.}
\label{guinierfit}
\end{figure}

Figure~\ref{guinierfit} features the magnetic Guinier analysis on nanocrystalline Co. Figure~\ref{guinierfit}(a) shows the Guinier plots, i.e., $\ln[d \Sigma_{SM} / d \Omega]$ \textit{vs.} $q^2$, along with the weighted linear least-squares fits to Eq.~(\ref{sigmamtaylorall}), whereas Fig.~\ref{guinierfit}(b) displays the obtained $R^2_{GSM}$ as a function of $H_i^{-1}$ together with a weighted linear least-squares fit to Eq.~(\ref{refftaylorpara}). In Fig.~\ref{guinierfit}(c) the field dependence of $\frac{d \Sigma_{SM}}{d \Omega}(q = 0)$ is displayed. The Guinier plots in Fig.~\ref{guinierfit}(a) reveal that straight-line fits may not be appropriate for the data at the two smallest internal fields of $0.213 \, \mathrm{T}$ and $0.252 \, \mathrm{T}$, where an upward curvature becomes visible at the smallest $q$, in contrast to the data at higher field. In line with this observation we see that the data set in Fig.~\ref{guinierfit}(c) starts to deviate from the expected linear behavior for these two smallest internal fields (open symbols). This discrepancy can be explained with growing deviations from the small-misalignment approximation for decreasing fields, and can be taken as a criterion for the validity range of the approach. Therefore, the two data points at $0.213 \, \mathrm{T}$ and $0.252 \, \mathrm{T}$ were not taken into account in the Guinier analysis, which yields $R_{GH} = 20.5 \pm 1.2 \, \mathrm{nm}$ and $A = (1.5 \pm 0.2) \times 10^{-11} \, \mathrm{J/m}$ [Fig.~\ref{guinierfit}(b)]. The $A$-value perfectly fits within the range of values reported in the literature~\cite{kronfahn03,skomskibook}, while the $R_{GH}$-value corresponds to a spherical particle radius of $R \cong 26.5 \, \mathrm{nm}$, assuming the relation $R_{GH}^2 = \frac{3}{5} R^2$, which is valid for monodisperse particles. This value is larger than the average crystallite size of $10 \, \mathrm{nm}$ (determined by X-ray diffraction), an observation, which can be naturally explained by the presence of a particle-size distribution in our Co sample. It is well-known from nuclear SANS theory that a size distribution strongly weighs the $R_G$-value towards the largest features in the distribution; for instance, for spherical particles and point collimation, $R^2_G$ is then related to the ratio of the eighth over the sixth moment of the size distribution~\cite{kostorz,feigin}. Therefore, for the determination of the scaling relation between $R_{GH}$ and the average crystallite size, knowledge on the particle-size distribution is required. Lastly, as can be seen in Fig.~\ref{guinierfit}(c), the extrapolated forward-scattering cross section $\frac{d \Sigma_{SM}}{d \Omega}(q = 0)$ also obeys the theory prediction and follows the $\frac{d \Sigma_{SM}}{d \Omega}(q = 0) \propto H_i^{-2}$ scaling [compare Eq.~(\ref{rhtaylorpara})].

The present theory describes the magnetic SANS cross section, and its low-$q$ behavior (magnetic Guinier law), of polycrystalline bulk ferromagnets near magnetic saturation. It assumes that the perturbing magnetic anisotropy fields of the individual microstructural defects, which cause a perpendicular magnetization component and hence a contrast for magnetic SANS, vary randomly from defect site to defect site. For the particular case of a nanocrystalline bulk ferromagnet composed of single-crystal grains and atomically sharp grain boundaries (magnetocrystalline anisotropy only), the characteristic correlation length of the anisotropy-field variation is related to the average crystallite size. Other potential sources of spin inhomogeneity such as surface (grain-boundary) anisotropy or magnetoelastic anisotropy due to long-ranged stress fields are not explicitly included in our theory. Likewise, the approach is not expected to describe the magnetic SANS of \textit{inhomogeneously} magnetized nanoparticles, which are embedded in a nonmagnetic matrix. For such a microstructure, boundary conditions for the magnetization at the particle-matrix interface must be included into the micromagnetic description of SANS, for which there is currently no analytical solution. This poses a challenge for future studies.

\section{Conclusion}
\label{conclusion}

Based on the continuum theory of micromagnetics we have introduced the magnetic Guinier law for random-anisotropy-type ferromagnets [Eqs.~(\ref{sigmamtaylorall}) and (\ref{refftaylorpara})], and we have confirmed the validity of the approach by analyzing experimental data on nanocrystalline cobalt. The magnetic Guinier radius $R_{GSM}$ depends on both the nuclear grain (anisotropy-field) microstructure and on the magnetic interactions (exchange-stiffness constant, saturation magnetization, applied field). It can be quite generally determined from the analysis of the magnetic-field-dependent spin-misalignment SANS cross section, which is obtained by subtracting the nuclear and magnetic scattering in the saturated state from data at lower fields. The method is easily applicable to magnetic materials using unpolarized neutrons.

\section*{Acknowledgements}

Artem~Malyeyev acknowledges financial support from the National Research Fund of Luxembourg (AFR project No.~12417141). This work is based on experiments performed at the Institut Laue-Langevin, Grenoble, France. The cryomagnet used was funded by the UK EPSRC (Grant number:~EP/J016977/1). The nanocrystalline Co sample was kindly provided by Professor Uwe Erb from the University of Toronto.

\bibliographystyle{apsrev4-1}

\end{document}